\documentclass[compsoc]{IEEEtran}
\ifCLASSINFOpdf
   \usepackage[pdftex]{graphicx}
\else
\fi
\usepackage{csquotes}
\usepackage{eurosym}
\usepackage{hyperref}

\begin{document}
%
\title{Product Innovation through Internal Startup in Large Software Companies: a Case Study \thanks{* This is the authors’ version of the manuscript accepted for publication in the Proceedings of 42th Euromicro Conference on Software Engineering and Advanced Application (SEAA), 2016. Copyright owner's version can be accessed at \url{https://doi.org/10.1109/SEAA.2016.36}. This manuscript version is made available under the CC-BY-NC-ND 4.0 license. \url{http://creativecommons.org/licenses/by-nc-nd/4.0}} \thanks{** Please cite as: Henry Edison, Xiaofeng Wang, and Pekka Abrahamsson (2016). Product Innovation through Internal Startup in Large Software Companies: a Case Study. Proceedings of 42th Euromicro Conference on Software Engineering and Advanced Application (SEAA), 2016, 128--135.}}

\author{\IEEEauthorblockN{Henry Edison\IEEEauthorrefmark{1}\IEEEauthorrefmark{3},
Xiaofeng Wang\IEEEauthorrefmark{1}\IEEEauthorrefmark{3} and Pekka Abrahamsson\IEEEauthorrefmark{2}\IEEEauthorrefmark{3}}

\IEEEauthorblockA{\IEEEauthorrefmark{1}Free University of Bozen-Bolzano, Bolzano, 39100, Italy\\
\IEEEauthorrefmark{2}Norwegian University of Science Technology, Trondheim, NO-7491, Norway\\
\IEEEauthorrefmark{3}Software Startups Research Network, http://softwarestartups.org\\
Email: (henry.edison, xiaofeng.wang)@unibz.it, pekka.abrahamsson@id.ntnu.no}}


%


\maketitle

\begin{abstract}
Product innovation is a risky activity, but when successful, it enables large software companies accrue high profits and leapfrog the competition. Internal startups have been promoted as one way to foster product innovation in large companies, which allows them to innovate as startups do. However, internal startups in large companies are challenging endeavours despite of the promised benefits. How large software companies can leverage internal startups in software product innovation is not fully understood due to the scarcity of the relevant studies. Based on a conceptual framework that combines the elements from the Lean startup approach and an internal corporate venturing model, we conducted a case study of a large software company to examine how a new product was developed through the internal startup effort and struggled to achieve the desired outcomes set by the management. As a result, the conceptual framework was further developed into a Lean startup-enabled new product development model for large software companies.
\end{abstract}


\section{Introduction}
It is widely accepted that product innovation is vital to companies to sustain their competitive advantages (e.g. \cite{teece07,chandy00,kuratko14}). Product innovation refers to the creation and introduction of new (technologically new or significantly improved) products which are different from existing products \cite{hull10,oecd05,hage99}. Through product innovation, companies are able to create new market and entry barriers, challenge market leaders and leapfrog competition \cite{brown95}. Companies are able to accrue high profit because at the time a new product is released, there is no competition in the market until imitators produce similar products \cite{roberts99}.

Developing product innovation is a risky activity \cite{kleinschmidt91,song98}. Many companies are too risk-averse to engage in any innovation initiatives \cite{ahmed98}. As in automated factories, people in large companies are trained to do prescribed and specific tasks reliably. Hence, any endeavour to change the status-quo will emerge resistance. The implementation of an innovative idea must compete with other product development activities \cite{deVen86,connor13}. This is the case for software industry, too. Little attention is given to product innovativeness \cite{rifkin01}.

The studies on corporate entrepreneurship suggest that internal startup is an ideal environment to nurture innovation and entrepreneurship in large companies \cite{kuratko09}. Although the internal startup is still operating within the corporation, the way of working is different with respect to the traditional research and development (R\&D) system. An internal startup takes the responsibility from end to end; from finding a business idea to developing a new product and introducing it to market \cite{bart88}. Therefore, internal startup is also seen as a learning process to create new competence different from the main business. Competence makes a difference among companies in yielding the outputs.

Even though an increasing number of software companies have adopted internal startup for the purpose of product innovation \cite{birkinshaw05}, practicing startup initiative within large companies can be a challenging endeavour \cite{breuer13}. Very few studies have investigated how large software companies can utilise internal startup to improve their competence and capability of product innovation. Based on this observation, the research question investigated in this study is:	\textit{``How do large companies leverage internal startup in software product innovation?}

To answer the research question, a conceptual framework based on the Lean startup approach \cite{ries11} and an internal corporate venture model \cite{burgelman83} is constructed. The framework enables a systematic analysis of a case study conducted in a large software company, to better understand the key processes of new product development through an internal startup in large software companies.

The remainder of this paper is structured as follows. The related work is described in Section  \ref{sec:literature_review}. Section \ref{sec:theoretical_development} presents the conceptual framework used in this study. Section \ref{sec:research_methodology} introduces the case studied and describes the research approach followed. Section \ref{sec:findings} presents the results, which are further discussed in Section \ref{sec:discussion}. Section \ref{sec:conclusion} concludes the paper with the summary of the major findings and an outline of future work.

\section{Related work}
\label{sec:literature_review}

\subsection{Software product innovation}
Like any modern dynamic business, software industry is highly influenced by its knowledge intensive and technology driven nature \cite{nambisan02}. Continual reliance on old or existing technology will jeopardise the market position of a company \cite{dibrell08}. Hence, companies must seek innovation as it disrupts former key players and creates new business or markets. Innovation is an iterative process that starts with identification of new market needs, which leads to the development of a solution that fits to that needs \cite{garcia02,drucker85}.

Software product innovation is concerned with introducing new software product to an existing or new market \cite{lippoldt09}. In software industry, product innovation occurs in either software or hardware development, or both, which raises strategic challenges for software companies \cite{kaltenecker15}. New companies emerge with innovative solutions. Some established companies manage to respond and survive, e.g., Google and Samsung in the mobile device and service domain, but others lose their business e.g. Nokia. Another example is open source software (OSS) \cite{bonaccorsi06}. Through OSS, many startups are able to enter a market and become a threat to the market leaders e.g. Linux against Microsoft Windows, Mozilla against Microsoft Internet Explorer, etc.

In the context of innovation in large software companies, Misra et al. \cite{misra05} develop a goal-driven measurement framework to measure innovation activities in companies. The framework adopted the Goal-Question-Metric (GQM) approach to define the goals of innovation program and the metrics to measure their achievement. Although they provided a set of metrics for measuring innovation, the study did not present a clear methodology on how they defined the goal, questions and metrics. Nor did it explain clearly the relationship between the suggested metrics and innovation. In addition, it is still unclear what innovation activities constitute. A study of Gorschek et al. \cite{gorschek10} proposes a model for an early stage innovation, which emphasises on ideation and selection prior to actual development. However, the understanding of the end-to-end development from ideation to commercialisation is yet to be fully achieved.

\subsection{Internal startup}
The literature of corporate entrepreneurship shows that to generate revenue streams and values for shareholders, large companies are engaged in either internal startup (or internal corporate venture \cite{huumo15}) or external startup (or external corporate venture) \cite{narayanan09}, or both. external startup usually includes joint venturing (e.g. Sony Ericsson), acquisition (e.g. the acquisition of Skype by Microsoft or Whatsapp by Facebook) and corporate venture capital (e.g. Google Venture, Intel Capital). On the contrary, in the case of internal startup, innovation is generated through a separate and dedicated entity which is operated within an established company and using resources that are solely under the control of the company \cite{narayanan09,roberts85}.   

Different models have been developed to explain the actual process of internal startup in companies. Burgelman \cite{burgelman83} identifies four major processes in a process model of internal startup: definition, impetus, strategic context and structural context. There are also three roles involved in the processes: corporate management, new venture division (NVD) management and venture lead (or intrapreneur). The definition process occurs in the R\&D department which is responsible to generate ideas for new business. The new idea combines the available technology and the market needs which should not be in line with the current corporate strategy. In this process, NVD plays an important role to coach the intrapreneurs in developing new business. To move from definition to impetus process, a product champion is required to mobilise resources needed. The impetus process refers to the entrepreneurial and organisational stages of development. It involves two stages: strategic forcing and strategic building. In strategic forcing, intrapreneurs are focusing on commercialisation of the new product. For the continuation of the impetus process, strategic building process takes place. Through strategic building activities, intrapreneurs are able to overcome the limitations of one-product and maintain growth rate, as required by the management. To establish its sustainability, the new business must get support strategically. Since any startup efforts may emerge from the bottom, corporate management applies selecting mechanism to ensure that only the ones that show the potential for fast growth have more opportunities to survive. Therefore, in the strategy context, intrapreneurs attempt to convince corporate management that the new business area can be part of current corporate strategy. Even when the current strategy cannot accommodate it, they must convince them to extend the strategy to protect the initiative. Therefore, an organisational champion plays an important role to communicate with the management about the new business area. 

The study by Garud and Van de Ven \cite{garud92} develops a model based on trial-and-error learning to overcome the uncertainty and ambiguity of the internal startup process. Before embarking on next activities, an intrapreneur evaluates the outcomes of prior activities. If the outcome is positive, the intrapreneur proceeds to the next activities, otherwise a change to plan is needed. A champion from corporation is needed when a series of negative outcomes occur or major environmental changes happen. The plan is reviewed to seek for alternative activities. Thus, the champion serves as a mentor to guide the changes in the plan. 

A recent internal startup model was introduced by Breuer \cite{breuer13}. She argues that there are five activities in specifying business model for a new venture: exploration, elaboration, evaluation, experimentation and evolution. However, the framework assumes that intrapreneurs already have identified the values that will be delivered to the customers. In reality, it is the customer values that intrapreneurs must seek and validate first \cite{blank05}.

In summary, the current internal startup process models focus more on the resource-based view of the company. The model from Burgelman \cite{burgelman83} acknowledges that the idea of new business is the combination of technology and market needs. However, the model does not give any clue how this can achieve the problem/solution fit. Also it does not recognise the learning process in developing a new product. Both models from \cite{garud92} and \cite{breuer13} are taking into account the learning process to create knowledge. They are able to describe the dynamics of creating new business but fail to explain the interaction between the internal startup and the parent company.

\section{Conceptual Framework}
\label{sec:theoretical_development}
To guide the study process to answer our research question, we drew upon the Lean startup approach, which has been suggested by \cite{ries11}. It is a hypothesis-driven approach \cite{eisenmann12} which aims at achieving the problem/solution fit first and then the product/market fit. To capture customer value, an entrepreneur start a feedback loop that turns an idea into a product then learn whether the business hypotheses are valid or not. This can be done by developing a minimum viable product (MVP) using an agile method and collecting customer feedback on the product. The feedback becomes the input to validate the hypotheses and improve the product. As the result, the startup might pursue a new direction of the business, called pivot in the Lean startup term, or continue and scale the proven business model. Pivot is common to any startup, since it could prevent the startup from bankruptcy if time between pivots is minimised. Even though originally conceived for standalone startups, the Lean startup approach was claimed to be useful in large corporate settings as well \cite{ries11}. 

To guide the empirical study, we constructed a conceptual framework by bringing together the elements from Burgelman's internal startup process model \cite{burgelman83} and the Lean startup approach \cite{eisenmann12}. Table \ref{tab:conceptual_framework} shows the major processes and their related activities. In this framework, the innovation initiative can be either top down (top management driven) or bottom up (employee driven). 

\begin{table*}[htbp]
    \caption{Conceptual framework}
    \label{tab:conceptual_framework}
    \begin{tabular}{p{1.5cm}p{3cm}p{12.2cm}}
    \hline \hline
    Major process & Description & Key activities\\
    \hline
    Vision & Setting up the vision & \emph{Envision}: Adopting the Lean startup model, the innovation initiative starts with envision, where intrapreneurs set the vision and translates the vision into hypotheses. To do this, they need two things: \emph{authorisation} from corporate management and \emph{coaches} from NVD management on how to turn the vision into successful business.\\
   Steer & Validating hypotheses & The initiative needs a product champion to get further resources. Once it gets approval from top management, the \emph{build-measure-learn loop} takes place to validate all hypotheses. Base on the learning, intrapreneurs have two options: \emph{pivot or persevere (then scale)}. When all the hypotheses are valid, then it is the time to integrate the new business into the company portfolio. \\
   Accelerate & Leveraging the new product into strategic context & In this process, there are two activities: \emph{selection and rationalising}. To control innovation initiatives in the company, corporate management uses selecting mechanism. Only the initiatives that have greater potential impact get continuous support. In rationalising process, the intrapreneurs must convince corporate management to change the strategy to accommodate the new business. In parallel, the NVD management plays an important role to \emph{delineate} the new business in the current strategy. Therefore, organisational championship is needed to continuously communicate with corporate management about the development of new business area. \\
   \hline \hline
   \end{tabular} 
\end{table*}

\section{Research Approach}
\label{sec:research_methodology}
The present study aims at developing a good understanding on how large software companies leverage internal startup in their software product innovation. Due to the complex nature of the research phenomenon and the intention to achieve an in-depth understanding of it, a single case study \cite{runeson12} is considered a suitable research approach. The criteria to select a case company are: (1) the company develops software in-house, (2) the company has at least one dedicated team that is responsible from the ideation to commercialisation of a new software; and (3) the area of the new software product falls out of the current main product line. The unit of analysis in this study is a development team of new software.

\subsection{Case company and context}
The company under study is F-Secure, a large Finnish cyber security and privacy company. F-Secure was established in 1998. Currently, F-Secure has more than 900 employees in 25 countries around the globe. Its products are available from over 6000 resellers and 200 operators in more than 40 countries. Tens of millions of users all over the world are using its products. In 2014, F-Secure generates revenue of 137 million and earnings before interest and taxes (EBIT) of 23.3 million. 

\begin{figure*}[htbp]
    \centering
    \includegraphics[width=\textwidth]{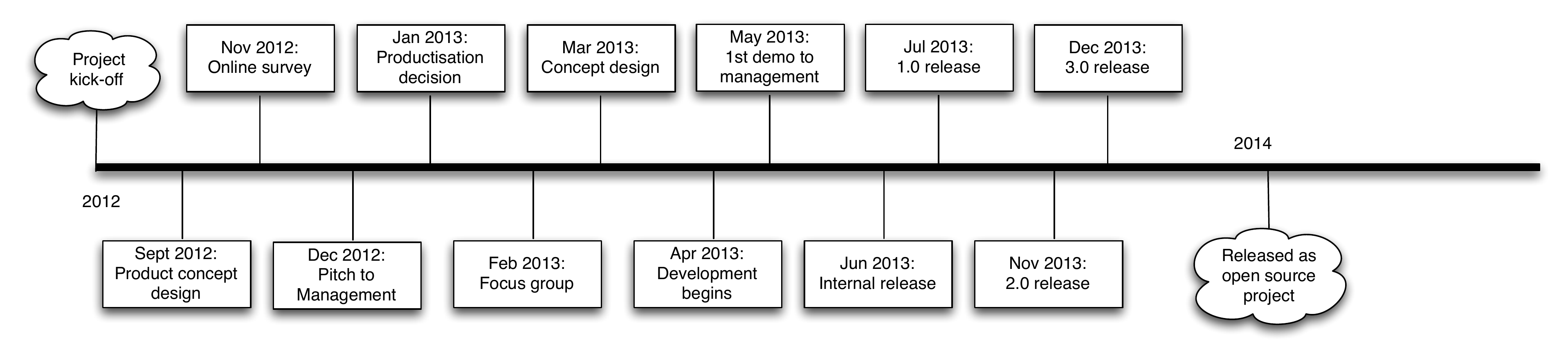}
    \caption{Project timeline}
    \label{fig:timeline}
\end{figure*}

Over the years, F-Secure has a long experience in various internal innovation e.g. ground up innovation, 10\% free time, hackathons. In 2012, F-Secure was running a growth strategy exercise to explore a new opportunity area, which was people protection. An internal startup team was established to find a concrete idea for new software product around people protection, called Lokki. A concept design was pitched to corporate management in December 2012. Lokki was released in July 2013 publicly. The timeline of the product development is shown in Fig. \ref{fig:timeline}. In Summer 2013, the corporate management decided to change the company's strategy. Lokki was no longer within the scope of the strategy. F-Secure decided to continue the development as an open source project, collaborating with leading universities in Europe and US and Facebook.. Based on the criteria described previously, F-Secure is a suitable case for this study.

\subsection{Data collection and analysis}
The conceptual framework presented in Table \ref{tab:conceptual_framework} serves as the theoretical lens for the investigation of the case, acting as a sensitising and sense-making device that guides the data collection and analysis processes. It was used to frame the interview questions, and enabled a holistic understanding of the dynamics between the internal startup and other entities inside the company. As a result, the framework was instantiated, modified or extended to better explain the empirical observations.

Semi-structured interviews were used as the primary data collection method \cite{robson11}. To better understand the phenomenon, eight employees with different roles in F-Secure were interviewed (see Table \ref{tab:interviewees_information}). Each interview lasted approximately 45-60 minutes. All interviews were transcribed verbatim. Field notes were taken during the interviews. Other supporting materials, such as internal corporate documents, presentations, white papers, etc., were also collected, to help achieve a more comprehensive understanding of the case.

\begin{table}[htbp]
    \caption{Background information of interviewees}
    \label{tab:interviewees_information}
    \begin{tabular}{p{2.5cm}p{2cm}p{3cm}}
    \hline \hline
    Role & Years of experience at F-Secure & Role and responsibility\\
    \hline
    Chief Information Officer (CIO) & $>$15  & Part of leadership team, heading R\&D operation \\ 
    Team Lead & $>$5  & Leading the internal startup team \\
    Director & $>$6 & Heading innovation operation\\
    Senior User Experience Designer & $>$4  & Conceptualising the applications and creating visual user experience design for mobile and desktop application \\
    Senior Manager & $>$8  & Product line manager and software architect \\
    Manager & $>$5 & Managing ground-up innovation \\
    Lead Software Engineer & $>$7  & Leading software development team \\
    Software Engineer & $>$5 & Server-side software developer\\
   \hline \hline
   \end{tabular} 
\end{table}

The data analysis process was conducted iteratively as encouraged by the grounded theory methodology \cite{charnaz06}. The interview data was coded line by line with the priori codes \cite{saldana12}, which focused the analysis around the innovation activities and their impacts on the internal startup team and the company. The documents obtained from the interviewees and field notes were also included in the coding steps, which allowed us to triangulate the interview data. During the axial coding, we sorted the coded materials into key processes as described by the framework, the actors who performed the activities and identified the factors that linked between two processes. The initial codes used during the analysis and coding phase were derived from the conceptual framework as shown in Table \ref{tab:conceptual_framework}.

\section{Findings}
\label{sec:findings}
As shown in the timeline, the team did two iterations in their development process. The first iteration started with the product concept design in September 2012 until they conducted the focus group. After the focus group they learned that the customers wanted Lokki. Thus, the team switched to Lokki and went back to concept design in March 2013, and went through another iteration of product development. In the following sub-sections the findings of this study were presented in detail, structured according to the conceptual framework (as shown in Table \ref{tab:conceptual_framework}): Vision, Steer and Accelerate. 

\subsection{Vision}
\label{sec:vision}
One senior employee (who later became the team lead) was assigned to prepare the concept creation within the people protection area, and to build the internal startup team. The team lead wanted to have all the competences in the team, e.g. developers, user experience designers, marketing, tester, etc. Therefore, he did one-to-one interviews to recruit the team members internally. It actually took 2 months to find the members with entrepreneurial mindset. It was not easy to build an entrepreneurial team from the existing pool of employees, because:
\begin{displayquote}
\textit{``They all wanted to hear that nothing is gonna happen and we're gonna be great for years. The problem is I cannot tell that. I mean, I don't know what's gonna happening.''} [Senior Manager]
\end{displayquote}

The internal startup team were dedicated to this project only. All the team members still got their monthly salary. The team had as least dependencies to the corporate as possible to shorten development cycle and reduce time-to-market. It is different from a R\&D unit since the internal startup operates only in the scope defined for them:
\begin{displayquote}
\textit{``[The startup is] controlled, because the focus is innovation. ... the business case is approved, and they are running towards the target under that business case approval and making experiments, maybe even pivot.''} [CIO]
\end{displayquote}

The team got a mandate from management to seek seeds for new product around people protection area, something that the company did not have yet. The team developed some scenarios of potential problems and identified the potential solutions. Eventually, as the team lead described, after \textit{``stealing some hour from some people in the company who get interested in this idea and ... working some afternoons, some evenings''}, the team landed with the idea of family location sharing. Using the family location sharing application, parents are able to see the locations of their kids or other family members. Therefore, if something bad happens to their kids, they can easily locate the position and help them.

The new idea was validated through an online survey which was published globally, to see if it resonates in the best way with the users. The questionnaire results showed that there was a big interest for the potential users to have a software that allows them to know where their family members or friends are. The new idea was then pitched to corporate management on December 2012. However, the new idea was questioned by the product management who was afraid of cannibalising:
\begin{displayquote}
\textit{``Then the first question from product management will be, ok, are we cannibalising anything? ... They do not want to harm the business which [has been] successful ...''} [Senior Manager]
\end{displayquote}

At that moment, it was still unclear to the team what the value proposition of the new product was. However, there was one member of the top management supported this concept. The management liked the concept and authorised the team to start developing the new product for several platforms.  

The role of NVD management to coach the team was not recognised in the case company. It was the first time for the company to work in a startup manner to develop new product. Nobody in the company knew how it should work:
\begin{displayquote}
\textit{``Yes, I was told to work like a lean start-up, ... than I was discussing this with [the top management] and `What do you actually mean by that?' He doesn't really know, but [the lead's name] go and find out...''} [Team Lead]
\end{displayquote} 

In the case of F-Secure, the corporate management took the responsibility of product championing because they were deeply involved in the vision process. Hence, the team did not have any difficulties in resource procurement e.g. people, budget, infrastructure etc., which were already available in the company. In addition, they got backup from one member of corporate management who supported the new idea. 

\subsection{Steer}
\label{sec:steer}
Using flip-chart prototyping, the team developed three different product concepts. Six families with their kids from internal employees were invited to a focus group to evaluate those concepts. It was surprising for the team to find out that the concept that they thought were be the best was not selected by the participants:
\begin{displayquote}
\textit{``We were sure that they would select this one [the other product] but they select this one [Lokki]... They selected not the one which we internally hoped, wanted for.''} [Lead Software Engineer]
\end{displayquote}

Since the team was responsible for the whole processes, they managed to have close communication with the users. They got 650 emails from the users asking for new features or improvement. This enabled them to identify which features brought value to the users, and hence needed to be implemented. The request for new features also came from the management or the team itself, but not all of them were put into the backlog. It was the team lead's decision as to which feature would be implemented in certain release.

During the development process, the team broke some rules in the company. An example of this was the team wanted to launch the product in Russia. They needed the localisation ready on a certain date. If they followed the standard process, they would not be able to meet the deadline. Hence, one team member who spoke Russian as his mother tongue did the localisation by translating all the marketing texts by himself. This unusual way of working enabled the team to reduce the development time cycle, but raised an internal conflict inside the company because some people felt that the team had trespassed their territory:
\begin{displayquote}
\textit{``... a formal document that writes 'your title is product marketing manager and your job is creating something. Your responsibilities are this, this, and this. And you report to this person'. So, this person sends me that document that had been defined for her when she had a job in the company. And then she said 'hey [Team lead's name] you are doing in your team the tasks that belong to me'. And I said 'yeah, but you choose not [to do the task]', and she was really really angry.''}[Team Lead]
\end{displayquote}

The MVP was demoed to the management in May 2013. It had three main features: invite other users to the close group, being able to see them in the map, and being able to show/hide the location. Then the team did a survey in some schools to test the product and gather feedback from the parents and children. The first version was released for iTunes and internal users in June 2013, and then for Google Play in August 2013, and finally for Windows Store in January 2014. The team internally set to get 30,000 users by Summer 2014. The team did not have enough budget in marketing, hence they used their personal network to attract new users.
 
One of the existing location-based product on the market at that moment was Google Latitude, which also allowed users to see their locations on the map. However, Google discontinued that service in 2013, hence Lokki got a lot of users who switched from Latitude. To get more users, the team intended to enter the US market. However it was not easy to sell the product related to kids in US because there was an online privacy protection act. The team needed legal experts to find a way to enter the US market:
\begin{displayquote}
\textit{``We had dialogue with our in house lawyers and then they found a legal consultant who advised to ask for approval from user. So when I decide to use Lokki, it will ask me that `Hey, am I under, 12 or 15?' and if I answer under that age, it will tell me that `Hey, do you have permission from your guardians/parents for you to use this?' and then you can check a tip box, `Yes I have permission.' And they were happy with this approach.}'' [Team Lead]
\end{displayquote}

During the steer process, the team was monitored and evaluated by the management in every three months. The first evaluation was done in May 2013, when the team did the first demo to the management. In this evaluation, the management wanted to see whether the team was able to develop new product in a short cycle. In the past, the company has one product per year with maybe two releases. The internal startup team managed to create product from the scratch and to release it on several platforms in three months. Due to this success, all the team members got a bonus of 2-3 months' salary on top of their normal salary.

The management used two measures to evaluate the team progress: number of users and user satisfaction. Number of users was based on the number of downloads from the apps store e.g. Google Play or iTunes store. User satisfaction was measured by NPS (Net Promotor Score). NPS basically asks one question from the end users: whether someone would promote the product to his friends or colleagues. The answer is based on 0 to 10 scale. 

In November 2013, there was a debate at the corporate management level. There was no agreement on the continuation of this product. Some of them wanted the team to develop premium features, which was expected to bring financial return to the company but others argued the other way around: number of users first, then money. It was unclear for the team what would be the target that needed to be achieved:
\begin{displayquote}
\textit{``[It was] confusing even slightly demotivational also.''} [Team Lead]
\end{displayquote}

The role of organisational champion was not recognised in the case company during the steering process. When the strategy changed, there was no one from corporate management who protected the initiative. Moreover, there was a change in the management that put team to a halt:
\begin{displayquote}
\textit{``So in a way we were, for three months we were in a limbo, we were not even [receiving] high quality guidance from the leadership team.''} [Team Lead]
\end{displayquote}

\subsection{Accelerate}
\label{sec:srs}
In Summer 2013, the company's strategy was updated. The people protection area was no longer inside the boundary of the strategy. A new emphasis became more prominent in the company?s strategy. However, the team did not react to this situation. As described before, there was no organisational champion who could protect the team. This situation made it more difficult for the team to get additional support from the rest of the company:
\begin{displayquote} 
\textit{``We were not on their agenda at all. We were noise that was distracting.''} [Team Lead]
\end{displayquote}

While the corporate management seemed like the performance of the team and the product, the user growth was small. The reason was the need of such product in Europe was not big enough as compared to the US:
\begin{displayquote} 
\textit{``Because the target segment was kids who were starting schools and for the parents who could see where the kids were. And in [the country] ... is pretty safe ... and no accident so far, really... like one or two within ten years.''} [CIO]
\end{displayquote}

In early 2014, the corporate management decided to release Lokki as an open source software project collaborated with two universities in Europe and the US.

\section{Discussion}
\label{sec:discussion}
\begin{figure*}[htbp]
    \centering
    \includegraphics[width=\textwidth]{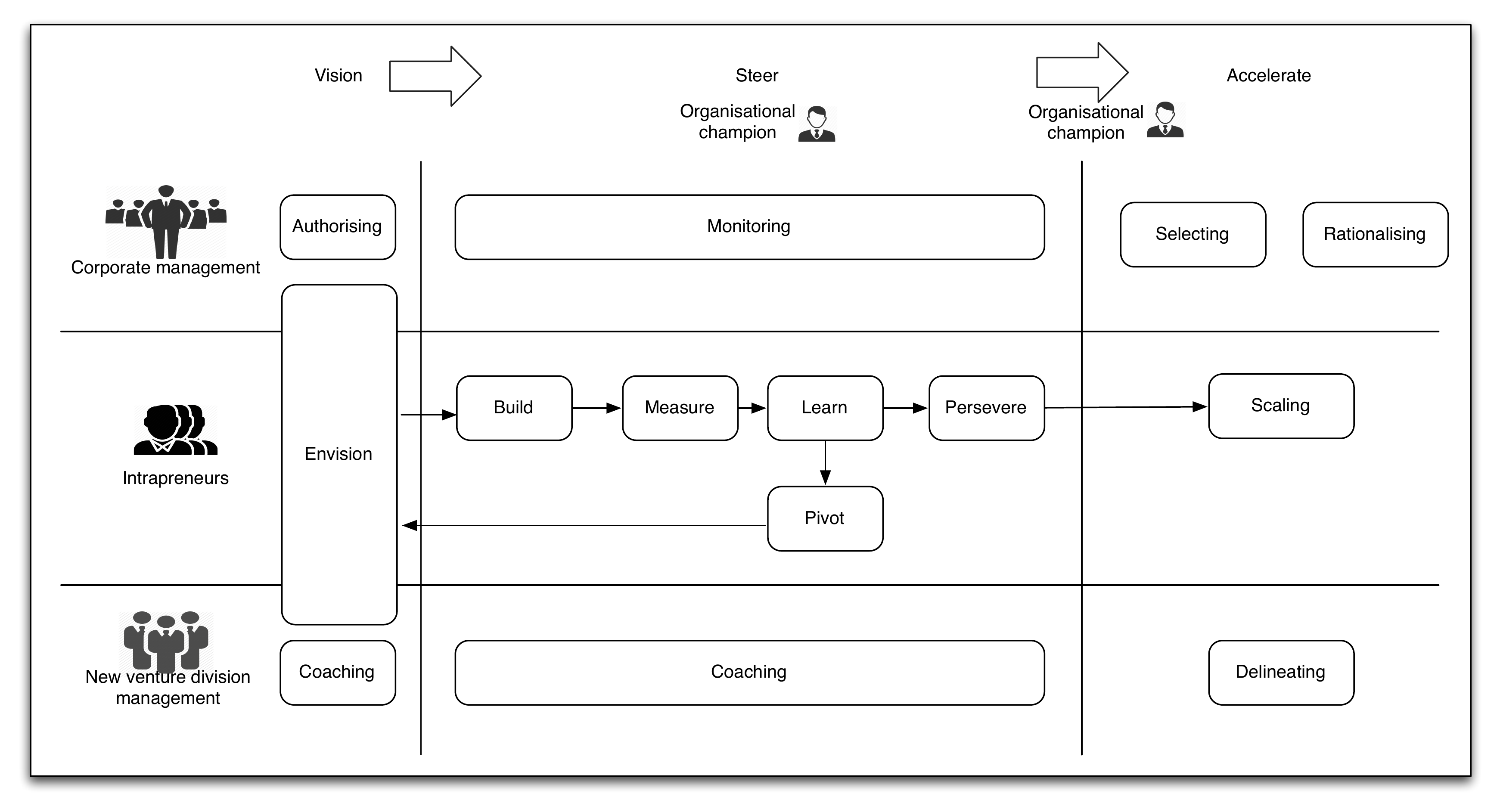}
    \caption{Lean startup-enabled new product development model}
    \label{fig:process_model}
\end{figure*}

The reflection on the findings obtained from the F-Secure case, in terms of what it has done well and should have implemented in its internal startup experiment, and the re-examination of the conceptual framework, led to a Lean startup-enabled new product development model, as shown in Fig. \ref{fig:process_model}.

As illustrated in Fig. \ref{fig:process_model}, the vision on the new product should be co-defined by the management and the internal startup team together. The management sets a boundary where the new idea should fall within, and the team explores the concrete idea. The new idea is then pitched to the management to get approval for development. While Burgelman \cite{burgelman83} suggests that the coaching from NVD is needed in the internal startup process, it was absent in the case of F-Secure. As described in Section \ref{sec:vision}, it was the first time for F-Secure to develop a new product in the internal startup manner. Therefore, there was no entity in the company that could coach the team during the whole project. If the NVD management existed, it should have been involved in the co-definition of the vision as well.

The case study findings show that working in a startup manner allows the team to easily communicate and talk with their customers, which helps them to identify what customers perceive value over the time. As a part of the corporation, the team could easily get support from the rest of the company. In the case of F-Secure, the internal startup team did not have to pay anything for legal consultant regarding the US market. Moreover, through the startup manner, the team is able to get out of the bureaucratic situation in creating the new product. However, this may raise the tension inside the company because sometimes they have to violate the standard operating procedures which have been the backbone to manage the corporation. 

To discover the real problem, Ries \cite{ries11} suggests to involve the customer since day one. In the F-Secure case, instead of going out of the building, the team made assumptions about the customer need by developing what they thought could be the real problem and then relied on the result of the focus group to test their assumptions (see Section \ref{sec:steer}). Later, they learned that the need of that product was not big enough to scale the production. The focus group approach has severe limitations \cite{ogawa06}. First, the results of the test do not reflect the reaction of broader population. Second, in focus group, customers are evaluating a verbal description of a concept, which might not communicate its true unique value proposition. Third, the results do not reveal the customers' real purchasing behaviour, only their intentions to purchase. The Lean startup principles of going out of the building and interact with real customers with MVP to obtain feedback would have been a better way for F-Secure to validate the product concept.

Ries \cite{ries11} states that validated learning is a process where entrepreneurs run experiment to validate their hypotheses about the customer needs to reveal current and future business prospects. Based on the results of hypotheses testing, they must decide whether to persevere with the original ideas or pivot to other directions. Ries \cite{ries11} identifies ten different types of pivot for startups. The reasons are related to product, business model and engine of growth. In the case of F-Secure, there were three pivots the internal startup project had or should have conducted. The first is the customer need pivot. As we can see in Section \ref{sec:steer}, the team made wrong assumption about Lokki. The second is the engine of growth pivot. After releasing the third version of the product, the team realised that the user growth was small. They learned that the need in the country where they are operating was not big enough. When they tried to penetrate to other countries, there were already competitors that were offering better service. The third pivot that the team should have done but did not was when the corporate strategy changed. By not reacting to it, it made the product development project out of the scope. As suggested by \cite{burgelman83}, in this situation the role of organisational champion is critical to convince top management either to change strategy to protect the idea or to map the business idea in the current strategy. This role is missing in the case of F-Secure.

Due to the single case study design, the findings of the study cannot be generalised to other software companies utilising internal startups for product innovation. Therefore, the framework needs to be refined and evaluated with more cases. The interviews could have been longer to allow a greater depth, even though they were conducted in short time to avoid the maturation threat \cite{wohlin00} and boredom effect \cite{juristo01} and to make the interviewees focus on the track.

\section{Conclusion and Future work}
\label{sec:conclusion}
Through software product innovation, large companies can keep their position one step ahead of their competitors, and startup initiative is considered a promising way to generate software product innovation in large companies. However, there is a lack of high quality studies on startup initiative inside large software companies.

In this study, we developed a Lean startup-enabled new product development model. The model describes the actual process of both product and business development. We identified the key activities of software product innovation through internal startup in large software companies. The model allows to take into consideration the learning opportunity in every activity. 

The contributions of this study are two-folds. First, for research community, our results contribute to an understanding of startup initiative inside large software companies to seek for innovation. Second, for practitioners, our model could be used as a benchmark to better understand and identify the opportunity to improve the current process of internal startup for product innovation.



We envision a future study which will investigate how effective internal startup can be for software product innovation and how it could be leveraged more effectively. Moreover a comparative study on different companies is needed as the basis to refine and evaluate the proposed framework.

\bibliographystyle{IEEEtran}
\bibliography{bare_conf}

\end{document}